# LLaMA-Gene: A General-purpose Gene Task Large Language Model Based on Instruction Fine-tuning

Wang Liang
*To whom correspondence should be addressed. E-mail:wangliang.f@gmail.com

**[Abstract]**Building a general-purpose task model similar to ChatGPT has been an important research direction for gene large language models. Instruction fine-tuning is a key component in building ChatGPT, but existing instructions are primarily based on natural language. Natural language and gene sequences have significant differences in tokenization and encoding. Therefore, constructing a multilingual model that can handle both natural language and gene sequences is crucial for solving this problem.In this paper, we expand the capabilities of the LLaMA large language model to include gene language. This involves expanding the vocabulary using the Byte Pair Encoding (BPE) method, specifically tailored for DNA and protein sequences, and conducting further pre-training on these sequences. We then convert various downstream gene task data into a unified format for instruction fine-tuning and further fine-tune the model on this data.Our study demonstrates that a mixed model of gene and natural language, fine-tuned with instructions, achieves results comparable to the current state-of-the-art (SOTA) in tasks such as gene classification and gene sequence interaction. This provides a promising direction for building a unified large language model for gene tasks.

# 1 introduction

Large language models have revolutionized the field of artificial intelligence. These models have also found significant applications in the analysis of DNA and protein sequences. For tasks centered around nucleic acids, models such as DNABert2, HyenaDNA, and ScBert have been developed. These studies primarily address challenges related to the classification and structure prediction of DNA sequences(*1~4*). Similarly, in the domain of protein-related tasks, models like ProTrans, ProteinBERT, and ESM2 have been introduced. These models excel in applications such as structure prediction and functional annotation of proteins(*5~9*).

However, current research on gene large models has not yet achieved the evolution from GPT to ChatGPT. Most studies still involve fine-tuning pre-trained models for different types of tasks, resulting in various sub-models to solve specific problems. Current gene large models are not yet capable of using a unified chat-based approach to address different types of tasks, as ChatGPT does.

Currently, some studies are exploring methods to construct a unified gene task model by introducing specific markers. DNAGPT adds more types of tokens to the pre-training input sequences to represent concepts such as "categories" and "numerical values." For instance, <A> represents a positive example, and <N> represents a negative example. ESM3 introduced 256 functional keyword tokens *(13)*, while LucaOne utilized protein annotation information. Studies like BioMedGPT, InstructProtein and ProtSt leverage large amounts of biological literature and annotated protein sequences for unified large model training and fine-tuning *(15,16)*.

These studies have achieved impressive results, but there is still a gap to bridge in building a unified large model for gene tasks.

**Challenge**: Firstly, in terms of sequence processing, these models employ different tokenization and encoding methods for various sequences. When designing downstream tasks, it is necessary to specifically consider these rules, making it difficult for different models to collaborate effectively. To distinguish between sequences, specific sequence markers are required, such as representing protein sequences as <protein>ACDEFGHIKLMNPQRSTVWY</protein>, and using different markers for DNA or structural data. There is also uncertainty about whether these specific markers will be recognized by large models.

Furthermore, different models have their own specific application ranges. For example, models like LucaOne and ESM3 primarily address downstream tasks such as gene sequence classification. BioMedGPT, on the other hand, focuses on biological question answering, literature summarization, and protein function prediction based on natural language. These studies have not yet reached the level of versatility seen in models like ChatGPT, which can handle a wide range of tasks including classification, summarization, reasoning, and computation.

Additionally, the parameter scales of current gene sequence models are generally smaller compared to mainstream large language models, typically below 1 billion parameters. This limitation makes it difficult for these models to exhibit emergent properties such as knowledge emergence and capability transfer. Currently, they are mostly limited to traditional sequence classification tasks and cannot fully leverage the advanced reasoning capabilities of larger models.

To address the above challenges, we first used the BPE (Byte Pair Encoding) method to perform uniform tokenization and encoding for natural language, DNA sequences, and protein sequences, without the need to introduce specific sequence markers. Then, we conducted continuous pre-training on the 7B-scale LLaMA large model using DNA and protein sequences. Further, we converted gene sequence-related classification and other data into unified instruction data and performed additional instruction tuning on this mixed-language model. The trained model demonstrates good performance in various downstream gene tasks.

# 2 Materials and methods

## 2.1 Overview of llama-gene

The llama-gene large model is built upon the LLaMA foundation and is obtained through continuous pre-training and instruction tuning using DNA and protein data.

Since the original llama model has been trained primarily on English natural language text, we expanded the vocabulary and conducted further pre-training using DNA and protein sequences. On this basis, we converted various gene sequence tasks, including protein and DNA classification tasks, protein-protein interaction tasks, and protein-DNA correlation tasks, into consistent instruction data using different prompt templates. We then used this instruction data to fine-tune the pre-trained model, resulting in the llama-gene model. The process is illustrated in the following diagram:

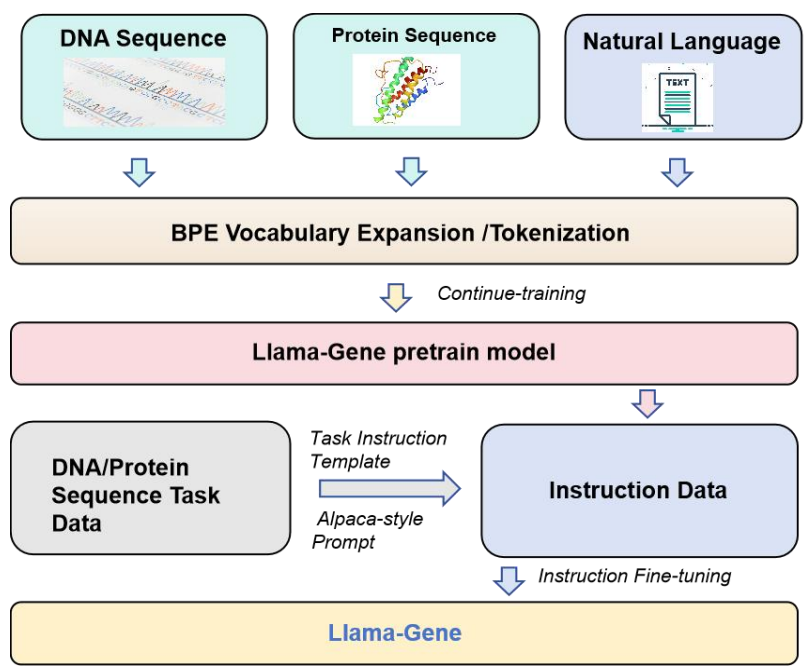

Fig.1    llama-gene training process. This model uses genome data, protein data and English research paper as continues pre-training corpora, employing the BPE method for uniform encoding without distinguishing between the three types of text. It then undergoes continues training on the llama network, resulting in a pre-trained model. Following this, specific prompt templates are used to convert gene downstream tasks into instructions. These instructions are then formatted into the standard Alpaca format, generating instruction data for fine-tuning the pre-trained model. This process leads to the creation of llama-gene. The usage of this model can adopt a dialogue mode based on prompt engineering.

## 2.2   Training datasets

The training data for LLaMA-Gene is roughly outlined in the table below:

Table.1 Training data

| period | datasets | data size | Data type |
| --- | --- | --- | --- |
| BPE training | multiple model organism genomes | 1G | DNA |
| BPE training | UniProt [multispecies] | 1G | Protein sequence |
| Continue training | multiple model organism genomes | 16G | DNA |
| Continue training | Swiss-Prot/TrEMBL | 16G | Protein sequence |
| Continue training | s2orc/biology paper | 16G | Natural Language |

| Instruction finetune | GLUE/convert | 100M | DNA downstream |
|---|---|---|---|
| Instruction finetune | lucaone/convert | 100M | Protein downstream |
| Instruction finetune | UniProt function | 300M | Protein translation |

### 2.2.1 BPE tokenlizer training datasets

For training the new BPE (Byte Pair Encoding) vocabulary, we use DNA sequences and protein sequences. The specific data used for this purpose is consistent with the data used during continuous pre-training

### 2.2.1 continues pre-training datasets

The training data for LLaMA-Gene is structured as follows:

**DNA Sequence Data:**We followed the pre-training data approach used by DNABert, extracting fragments of 300 to 1000 base pairs (bp) from multiple model organisms. The total data volume for DNA sequences is approximately 16 GB.

**Protein Sequence Data:**From the UniProt database, we extracted 16 GB of sequence data, including all data from Swiss-Prot and randomly selected data from TrEMBL.

**Natural Language Data:** During continuous pre-training, it is generally necessary to mix in some of the pre-training data to prevent catastrophic forgetting. We used biological research papers and other relevant texts as mixed corpora for this purpose.

### 2.2.2 Instruction Fine-tuning dataset

To adapt different downstream tasks into a unified instruction format, we construct standardized instruction data and then perform fine-tuning training. This method is also an important research approach in large natural language models, where the basic idea is to design prompt templates based on the characteristics of tasks such as classification, part-of-speech analysis, translation, etc., converting them into standard instruction fine-tuning data.

A typical example is NATURAL INSTRUCTIONS, a dataset developed by Allen AI Institute and other institutions, which includes 61 different NLP task datasets. The core concept of NATURAL INSTRUCTIONS is to express various NLP tasks as natural language instructions. Each task comes with detailed task descriptions, positive and negative examples, and relevant input-output pairs. This approach aims to enhance the model's ability to understand and execute diverse task instructions, enabling it to generalize better to new tasks. The dataset covers multiple task types, including text classification, question answering, summarization, text generation, and more.

We also follow this approach and use similar prompt templates to convert gene sequence-related downstream tasks into instruction data.

Fineture training data consists of instructions, inputs, and outputs. These three components are then formatted into a specific prompt format. Below are two common methods of formatting.Fig.5. In this article, we will use the Alpaca-style prompt formatting method.

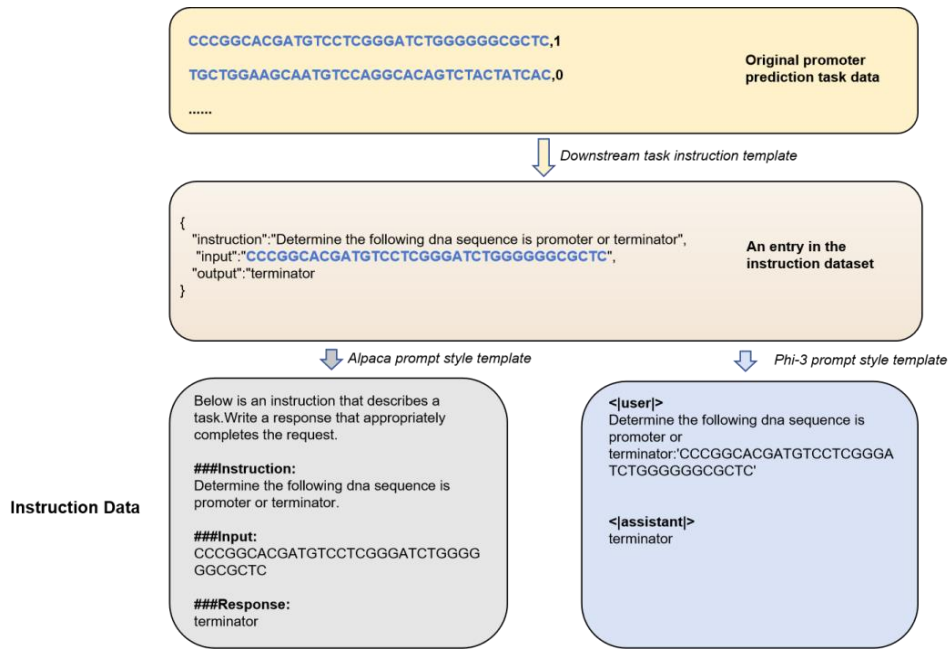

Fig.2 Build instruction datasets. For different downstream tasks, design specific instruction templates to convert downstream task data into instructions. Then, using the Alpaca prompt template, transform these instructions into uniformly formatted instruction fine-tuning data.

### 2.2.2.1 Classification task

For classification tasks, we have referenced templates used in natural language processing (NLP) to design our specific templates. The approach is detailed as follows:

Table.2 examples of a classification task

| Task name | sequence | sequence type | label name | label |
|-----------|----------|---------------|------------|-------|
| promoter detection | CATGCGGGTCGATAT... | DNA | Non-promoter | 0 |
| promoter detection | CTGAATCTCTCAGCC... | DNA | promoter | 1 |

For classification tasks, a template like the following can generally be used to construct instructions：

*{'instruction': 'Determine the {task_name} of following {sequence_type} sequence, The result will be one of the following: {label_name1}, {label_name2}, {...}',*

*'input': '.{sequence}',*

*'output': '{label_name}'}*

An example:

*{'instruction': 'Determine core promoter detection of following dna sequence, The result will be one of the following: Non-promoter, promoter.',*

*'input': 'CATGCGGGTCG...',*

*'output': 'Non-promoter'}*

## 2.2.2.2 Structure prediction task

For structure prediction tasks, we first convert them into general classification tasks and then adopt the classification prompt templates:

Table3. gene structure prediction task

| Task name | sequence | sequence type | secondary structure |
|---|---|---|---|
| Secondary structure prediction | MKQDKGLFDLVT | protein | HHHCCCCEEECCC |

To predict the secondary structure of a token, we use a sequence that includes this token and the preceding 30 base pairs (bp) as input. The output is the secondary structure of the last token in this sequence. Since the tokenization of sequences does not perfectly correspond to their secondary structures, it is necessary to align gene tokens with secondary structure sequences and determine the type of secondary structure for each token. Given that one token may correspond to multiple secondary structures, we apply a simple rule: assign the most frequent secondary structure to the token. If there is a tie in frequency, we prioritize the structures in the order H (alpha helix), E (beta sheet), C (random coil).

Table.4 Gene structure prediction task

| Task name | sequence | sequence type | secondary structure |
|-----------|----------|---------------|---------------------|
| secondary structure prediction | MKQDKGLFDLVT | protein | random coil |

Template for structure prediction:

*{'instruction': 'Determine the {task_name} of following {sequence_type} sequence, The result will be one of the following: {structure_name1}, {structure_name2}, {...}',*

*'input': '.{sequence}',*

*'output': '{structure_name}'}*

An example:

*{'instruction': 'Determine the secondary structure    prediction of following protein sequence, The result will be one of the following:    alpha helix,    beta sheet, random coil',*

*'input': '.MKQDKGLFDLVT',*

*'output': 'random coil'}*

## 2.2.2.3 Multiple sequence analysis task

For the multiple sequence association analysis task, we adopt the following prompt template to guide the model in identifying and analyzing associations between multiple sequences. This approach ensures that the model can effectively interpret the relationships and interactions among different gene sequences.

Table.5 multi sequence task

| Task name | sequence1 | sequence1 type | sequence2 | sequence 2 type | label | label name |
|-----------|-----------|----------------|-----------|-----------------|-------|------------|
| Conform to the Central Dogma | ACCAGTGCTCAGGT TAACAAAAT... | DNA | MAS GRL | protein | 1 | Not Conform |

| prediction | | | QLL AFAL .. | | | |
|---|---|---|---|---|---|---|

Template for multiple sequence:

{'instruction': 'Determine the {task_name} of following {sequence1_type} sequence and {sequence2_type} sequence , The result will be one of the following: {label_name1}, {label_name2}, ...',

'input': ' {sequence1_type} sequence: {sequence1} \n###\n{sequence2_type} sequence: {sequence2} ',

'output': '{label_name}'}

Example:

{'instruction': 'Determine the Conform to the Central Dogma prediction of following DNA sequence and Protein sequence , The result will be one of the following: Conform, Not Conform',

'input': ' DNA sequence: ACCAGTGCTCAG... \n###\n protein sequence: MASGRLQLLAFAL.. ',

'output': 'Not Conform'}

## 2.2.2.4 Function prediction

For the function prediction task, we have referenced translation templates used in natural language processing (NLP) to design our specific template. The approach is detailed as follows:

Table6. gene function prediction task

| Task name | sequence | sequence type | function |
|---|---|---|---|
| gene function prediction | GTTCTGCCCATAACA... | DNA | D-glucose transmembrane transport(GO:1904659) |

Template for function prediction:

*{'instruction': 'Determine {task_name} of following {sequence_type} sequence.',*

*'input': '{sequence}',*

*'output': '{function}'}*

An example:

*{'instruction': 'Determine gene function prediction of following dna sequence.',*

*'input': 'GTTCTGCCCATAACA...',*

*'output': 'D-glucose transmembrane transport(GO:1904659)'}*

## 2.2.2.5 Regression task

Large language models are generally not well-suited for handling regression tasks directly. Therefore, a common approach is to use a function call format to delegate the task to smaller, specialized models or scripts. Alternatively, regression problems can be approximated by converting them into classification tasks. For instance, you can transform a regression problem into a binary classification problem by categorizing values above the median as "high" and those below the median as "low." This method can also be extended to more categories if needed. Below is an example using binary classification:

Table7. gene regression task

| Task name | sequence | sequence type | value |
|-----------|----------|---------------|-------|
| gene expression level prediction | GTTCTGCCCATAACA... | DNA | 0.75 |

Since the regression predictions have already been normalized, we categorize the predictions as follows:

For values less than 0.5:      Label: 0    Name: low

For values greater than or equal to 0.5:      Label: 1      Name: high

Table8. gene function prediction task

| Task name | sequence | sequence type | label | label name |
|-----------|----------|---------------|-------|------------|
| gene expression level prediction | GTTCTGCCCATAACA... | DNA | 1 | high |

Template for Regression task:

*{'instruction': 'Determine the {task_name} of following {sequence_type} sequence, The result will be one of the following: {label_name1}, {label_name2}, {...}',*

*'input': '.{sequence}',*

*'output': '{label_name}'}*

Example:

*{'instruction': 'Determine the gene expression level prediction of following DNA sequence, The result will be one of the following:   low,   high',*

*'input': '.GTTCTGCCCATAACA...',*

*'output': 'high'}*

## 2.3 Training Strategy

### 2.3.1 Tokenization

We trained the vocabulary for DNA and protein sequences using BPE, then merged them into the original LLaMA vocabulary (32,000 tokens). Both the DNA and protein vocabularies consist of 30,000 words, resulting in a final vocabulary size of approximately 91,000 words.

To train this new mixed vocabulary, we adopted a two-step process to facilitate the validation of different sequence combinations' effects. First, we trained the DNA vocabulary and merged it into LLaMA. Then, we trained the protein vocabulary and added it to the model from the first step.

It's important to note that the training method used for LLaMA3 differs slightly from previous versions. Specifically, LLaMA3 uses a vocabulary trained with **tiktoken**, whereas earlier versions utilized the **SentencePiece** framework. When merging or working with these models, it is crucial to pay attention to the format of different frameworks.

### 2.3.2 Continue Pre-training

Due to the large number of parameters in LLaMA models, ranging from 7B to 405B, performing full-parameter continuous pre-training would be very costly. Therefore, we adopted the LoRA method for training, specifically using the PEFT framework from Hugging Face. The parameter settings were referenced from the configuration of chinese-llama.

The specific network layers for training are as follows:

**lora_trainable Parameters**

lora_trainable="q_proj,v_proj,k_proj,o_proj,gate_proj,down_proj,up_proj"

- **q_proj**: Query projection layer, used to generate query vectors in the attention mechanism.
- **v_proj**: Value projection layer, used to generate value vectors in the attention mechanism.
- **k_proj**: Key projection layer, used to generate key vectors in the attention mechanism.
- **o_proj**: Output projection layer, used to process the output after the attention mechanism.
- **gate_proj**: Gate projection layer, possibly used to control information flow, common in some variant models.
- **down_proj**: Down-projection layer, possibly used to reduce feature dimensions.
- **up_proj**: Up-projection layer, possibly used to increase feature dimensions.

These layers are primarily related to the attention mechanism and feed-forward networks. By specifying these layers as trainable, we can focus on adjusting the parameters of these critical parts during fine-tuning, thereby improving efficiency while maintaining model performance.

The network layers to be saved are as follows:

**modules_to_save Parameters**

modules_to_save="embed_tokens,lm_head"

- **embed_tokens**: Token embedding layer, responsible for converting input tokens into vector form.
- **lm_head**: Language model head, responsible for generating the final output from the hidden states of the model, such as predicting the probability distribution of the next word.

After fine-tuning, the states of these important modules can be correctly saved.

With this setup, approximately 10% of the parameters are being trained. On an 8-card L20 server, it takes about one week to complete training with 16GB of data.

**2.3.3 Instruction Fine-tuning**

The method for instruction tuning is entirely consistent with the training method of the llama pre-trained model. All use exactly the same causal language model head. This involves treating the constructed instruction data as general text sequences and inputting them into the llama model. In other words, instruction tuning is the same as the pre-training process, which is key to enabling multi-task handling. This is also key to how ChatGPT-like models are able to handle a variety of natural language tasks.

We typically train for 2 to 8 epochs. Since the training data is relatively small, fine-tuning on a L20 GPU*8 usually takes only about ten minutes to complete.

# 3 Experimental Results

## 3.1 Model Evaluation

Because we have essentially converted various downstream tasks into chat-based tasks. The assessment is based on whether the model's output tokens semantically match the expected ones. Therefore, We Use Accuracy (ACC) as the Primary Evaluation Metric.

For example, if the model outputs "promoter AGCC GGG" while the expected output is "promoter ", we still consider the model's prediction to be accurate. This is because, as a generative model, llama-gene is not specifically trained to recognize stop tokens. With more training data and a larger model size, it would be possible to produce outputs that exactly match the expected tokens.

## 3.2 Evaluation results

The evaluation datasets primarily referenced include those from DNABert2, lucaone, InstructionProtein, and BiomedGPT. These datasets encompass a variety of gene-related downstream tasks.

The specific evaluation results for LLaMA-Gene across these datasets demonstrate its performance on various gene-related downstream tasks. These results highlight the model's capabilities in handling diverse and complex gene-related challenges.The evaluation results for LLaMA-Gene are detailed as follows:

Table.9. Accuracy of different tasks

| Task | sequence type | llama-gene | SOTA |
|---|---|---|---|
| classification | DNA | 0.83 | 0.84 |
| classification | protein | 0.64 | 0.72 |

| structure prediction | DNA | 0.81 | 0.85 |
|---|---|---|---|
| structure prediction | protein | 0.73 | 0.85 |
| multiple sequence | DNA+DNA | 0.66 | 0.78 |
| multiple sequence | protein+protein | 0.63 | 0.87 |
| multiple sequence | DNA+protein | 0.71 | 0.91 |
| function prediction | DNA | 0.76 | 0.81 |
| function prediction | Protein | 0.71 | 0.81 |
| regression task | DNA | 0.83 | 0.87 |
| regression task | Protein | 0.82 | 0.86 |

The evaluation involves 2 to 3 specific datasets per task, primarily for validating the new method, which may not be universally applicable.

Compared to current SOTA methods, LLaMA-Gene performs adequately on DNA-related tasks but shows a noticeable gap in protein-related tasks. However, this result validates the new method's effectiveness. The performance is mainly limited by computational resources; LLaMA-Gene has been trained on relatively smaller datasets. Future research will use larger parameter scales and more extensive data to improve evaluation outcomes.

Currently, we use LLaMA base models from Generation 1 to 3, all with 7B or 8B parameters, showing no significant differences in gene-related tasks. This is because newer LLaMA versions focus on improving fine-tuning effects, which have not been applied to LLaMA-Gene. Future studies will explore using fine-tuned LLaMA models to transfer natural language inference capabilities to gene-related tasks.

Our instruction fine-tuning design offers versatility, allowing one model to handle various gene downstream tasks. Most existing large gene models require training different model heads for specific tasks, whereas our approach uses a generalized model.

More importantly, the instruction-tuned model can interact with users in a conversational manner and can leverage prompt engineering, RAG, chain of thought, and other techniques commonly used with large language models.

# 4 Conclusion

Advancing gene large models from the GPT era to the ChatGPT era is a current hot topic in the field of biology. This paper constructs the llama-gene model based on the LLaMA series of large models, validating the feasibility of a ChatGPT-style gene large model.

To address the challenges of building a unified gene task model, this paper introduces several innovations. First, a unified BPE tokenizer and encoding scheme are used for different types of sequences, significantly reducing the complexity of constructing and applying network models. Second, inspired by instruction construction methods in NLP research, we created instruction-tuning datasets for various downstream gene tasks. Based on these methods, we performed continuous pre-training and instruction tuning on multiple types of biological sequences using the LLaMA 7B model, ultimately constructing the llama-gene large model.

The llama-gene large model follows a similar construction pattern to ChatGPT, thus offering extensive application potential. For example, mature large model application frameworks like RAG (Retrieval-Augmented Generation) and Agent can be used to build comprehensive biological applications. These applications can integrate the strong logical reasoning capabilities of natural language with the understanding of biological sequences, enabling deep interpretation of biological sequences.